\documentclass[aps,twocolumn]{revtex4}
\usepackage{epsfig}
\usepackage{psfrag}

\begin{document}
\title{Comparing Simulation and Experiment of a 
       2D Granular Couette Shear Device}
\author{Marc L\"atzel, Stefan Luding\footnote{new address:
Particle Technology, DelftChemTech, TU Delft, Julianalaan 136,
2628 BL Delft, The Netherlands} and Hans~J. Herrmann}
\address{Institute for Computer Applications 1, Pfaffenwaldring 27, 70569
Stuttgart, Germany}
\author{Daniel W. Howell and R.~P. Behringer}
\address{Department of Physics and Center for Nonlinear and Complex
Systems, Duke University, Durham NC, 27708-0305, USA} 
\date{\today}

\begin{abstract} 
  We present experiments along with molecular dynamics (MD)
  simulations of a two-dimensional (2D) granular material in a Couette
  cell undergoing slow shearing.  The grains are disks confined
  between an inner, rotating wheel and a fixed outer ring. The
  simulation results are compared to experimental studies and
  quantitative agreement is found.  Tracking the positions and
  orientations of individual particles allows us to obtain density
  distributions, velocity and particle rotation rate for the system.
  The key issue of this paper is to show the extent to which {\em
  quantitative} agreement between an experiment and MD simulations is
  possible.  Besides many differences in model-details and the
  experiment, the qualitative features are nicely reproduced. We
  discuss the quantitative agreement/disagreement, give possible
  reasons, and outline further research perspectives.
  
\end{abstract}

\maketitle



\section{Introduction}


The pioneering work of Reynolds in 1885 \cite{reynolds85} and the more
elaborate investigations by Bagnold \cite{bagnold54} were among the
first experiments to closely address the problem of granular shearing.
Recently the subject of granular shearing has regained much interest
in the physics community due to the appearance of this process in
common granular flows such as convection \cite{knight97}, pipe and
chute flow \cite{drake90,nedderman80},
avalanches \cite{savage89,pouliquen96}, crack formation, and
earthquakes \cite{herrmann95}.

In the traditional picture for shearing of a dense granular material,
grains are assumed to be relatively hard so that they maintain their
volume and shape under applied forces.
If shear is applied to a granular sample, in principle, the grains
will respond elastically (i.e. reversibly) up to the
point of failure. The response in the elastic regime is still an
open question which is not addressed here \cite{vanel00,geng01,geng02}
because we focus on the regime of extended deformation.
Under shear, the grains will dilate
against a normal load, up to the point of failure.  
Under continued shearing the system appears to approach a steady
state, that is typically characterized by localized failure in
narrow regions
known as shear bands.  An extremely slow compaction/rearrangement
can also occur under steady shearing \cite{veje99}.
However, this effect will be disregarded in the following.
Here we are more concerned with the kinematics of the particles in
the ``quasi steady state''.


Recent experiments on granular shearing have primarily focused on the force
properties of the system 
\cite{miller96,veje98,veje99,howell99,howell99b,howell99c}.
Only a few experiments have explored the kinematics of shear zones,
and these involved using either inclined or vertical chutes
\cite{nedderman80,drake90,azanza99} or vibrated beds \cite{losert99}
where air flow between the particles may also have been important.
In a single case of
which we are aware, Buggisch and L\"offelmann \cite{buggisch89} 
investigated the mixing
mechanisms due to shearing in a 2D annular cell similar to the one
described here.  This experiment involved flexible boundaries, in contrast
to the fixed volume used in our work.


The simulations presented in this study are perhaps unique
in that they 1) match with considerable fidelity the parameters
in a set of experiments, and 2) the simulations and experiments
yield detailed properties which can be mutually compared.  Thus, the
goals of this work include the deeping of insight into an important
granular system, and the opportunity to explore how well a class
of models captures experimental observations.  The relevance of
this latter issue is underscored by a recent study in which diverse
groups modeling flow in a hopper, obtained an equally diverse
set of predictions, many of which did not match experiment 
\cite{rotter98,holst99,holst99b}.

In our experiment a 2D Couette cell filled with photoelastic polymer
disks is used to study both the mean and statistical properties of the
flow. Using particle tracking techniques, the spin and transport
velocity profiles as well as the associated density variations during
steady state shearing can be measured. Because the particles are
photoelastic, it is also possible to infer information on the local
stress state of the system, a topic which is considered elsewhere
\cite{geng02}.

The following are the key observations from these studies: A short
time after the beginning of shearing, a shear zone forms near the
inner wall.  The location of the shear band close to the inner wall
can be attributed to the fact that the shear stress is highest there
(decaying proportional to $(r-R)^{-2}$), as was observed from previous
simulations \cite{luding01,luding01b} and as is consistent with the static
equilibrium conditions of continuum theory in cylindrical coordinates.
The characteristic width of the induced shear band is found to be a
few particle diameters, almost independent of the average packing
fraction.  The mean azimuthal velocity decreases roughly exponentially
with distance from the inner shearing wheel, and within the
statistical fluctuations, there is shear rate invariance.  The mean
particle spin oscillates around zero as the distance, $r - R$ from the
wheel increases, but falls rapidly to zero away from the shearing
surface.  The distributions for the tangential velocity and the
particle spin show a complex shape particularly for the grain layer
nearest to the shearing surface, indicating a complicated dynamics,
where velocity distributions near the inner wheel are very wide and
non-Gaussian.

\subsection{Simulations}

There are a variety of numerical studies involving shearing.  Some of
these focus on on stress-strain relations
\cite{campbell85,evans83,walton86b,thornton91,thornton97b,thornton00b,latzel00,luding01,luding01b,masson01},
and others deal with shear banding in specific geometries
\cite{tillemans95,latzel00,luding01,luding01b,masson01,thornton01,oda00,oda98,radjai99}.
In the present study, we present MD simulations and investigate the
kinematic properties of a model system which was structured so that
its realization is as close as possible to the physical system
discussed here, a goal partially achieved already in
\cite{veje98,schollmann99}.

\subsection{Overview of the paper}

The rest of this paper is constructed as follows.  The experimental
set-up is reviewed in section \ref{sec:experiment}.  The similarities
and differences between the simulations and the physical system are
discussed in section \ref{sec:methods}.  The initial conditions and
the steady state are examined in section \ref{sec:initcond}, and the
results concerning velocity- and spin-distributions are presented in
section \ref{sec:diffpack} for different densities.


\section{Experimental Setup and Procedure}
\label{sec:experiment}

\begin{figure}
\epsfig{file=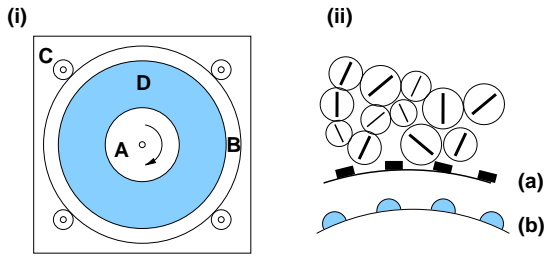,width=7cm}\\
\caption{{\bf (i)} schematic top view of the experimental setup.
{\bf (ii)} schematic drawing of the disks close to the shearing wheel. 
({\sf \bf a}) experimental realization of the walls. 
({\sf \bf b}) realization of the walls in the simulation.
\label{fig:apparatus}}
\end{figure}

In this section we give enough details so that the reader can
appreciate key aspects of the experiments and how the experiments
relate to the simulations.  The experimental setup and results are
discussed in more detail in Refs.\ \cite{veje98,howell99,howell99b}.
The apparatus, as sketched in Fig.~\ref{fig:apparatus}, consists of
(A) an inner shearing wheel (with radius $R$), and (B) an outer,
stationary ring confined by (C) planetary gears.  In the experiments,
a bimodal distribution of disks (D) is used, with about 400 larger
disks of diameter $0.899\,{\rm cm}$, and about 2500 smaller disks of
diameter $0.742\,{\rm cm}$.  An inhomogeneous distribution is useful,
since it limits the formation of hexagonally ordered regions over
large scales, even though there might still allow some short range
order \cite{luding02b}.  We use the diameter, $d=d_{\rm small}$, of
the smaller disks as a characteristic length scale throughout this
study.

The experimental walls are fixed, corresponding to a {\em constant volume 
boundary condition}.  All particles are inserted into the system
and the shear is applied via the inner wall for several rotations, before
averages in the nominally steady state are taken. If not explicitly 
mentioned, averages in the simulations are performed after about three 
rotations starting at $t=180$\,s, and extending over three rotations, 
until $t=360$\,s.

The mean packing fraction $\bar{\nu}$ (fractional area occupied by disks)
is varied in the experiment over the range $0.789\leq \bar{\nu} \leq
0.828$.  As we vary $\bar{\nu}$ we maintain the ratio of small to large
grains fixed, modulo small variations due to the fact that particle
numbers can only be adjusted by integer jumps.
Note that the effect of the wall particles
for the calculation of the global packing fraction is very small. 
For computing the packing fraction in the simulations,
only half the volume of the small particles glued to the side walls
is counted, so that these boundary particles always
contribute $\bar \nu_{\rm wall} = 0.0047$ to $\bar \nu$.

An important question is how the system response depends on
the shearing rate, which is set by $\Omega$, the rotation rate of
the inner wheel.  A variation of $\Omega$ over 
$0.0029\,{\rm s}^{-1} \leq \Omega \leq 0.09\,{\rm s}^{-1}$ in the
experiments shows rate independence in
the kinematic quantities, except for some small, apparently
non-systematic variations with $\Omega$. A few simulations with
$0.01$\,s$^{-1} \le \Omega \le 1.0$\,s$^{-1}$ showed
clear rate independence for the slower shearing rates 
$\Omega \le 0.1$\,s$^{-1}$, although the situation is less clear
at the higher end of these rates.

\section{Simulation Method and Similarity to the Experiment}
\label{sec:methods}

Details of the simulations have been presented elsewhere
\cite{latzel00}, and we will not repeat these.  However, we note that
the model is a soft-particle MD model. As noted, the parameters used
in the model were chosen to match the experiments as reasonably as
possible.  Specifically, the radii, static friction coefficient and
density of the particles, and the size of the container 
match the experimental values.  The boundary
conditions are chosen to mimic those in the experiment, see Sec.\
\ref{sec:experiment}. However, the ``teeth'' used on the inner and
outer ring of the experiment are replaced by small disks with diameter
$d_{\rm wall}$, see Fig.\ \ref{fig:apparatus}.  The properties of the
particles and the parameters for the (linear) force laws
\cite{latzel99,latzel00} are summarized in Table~\ref{tab:model_props}.
\begin{table}[htb]
\begin{center}
\begin{tabular}{|l||l|}
\hline
Property                              & Values\\
\hline
\hline
Diameter $d_{\rm small}$, Mass $m_{\rm small}$ 
                                      & $7.42$ mm, $0.275$ g\\
Diameter $d_{\rm large}$, Mass $m_{\rm large}$
                                      & $8.99$ mm, $0.490$ g\\
Wall-particle diameter $d_{\rm wall}$,& $2.50$ mm \\
System/disk-height $h$                & $6$ mm \\
Normal spring constant $k_n$          & $352.1$ N/m \\
Normal viscous coefficient $\gamma_n$ & $0.19$ kg/s \\
Tangential viscous damping $\gamma_t$ & $0.15$ kg/s \\
Coulomb friction coefficient $\mu$    & $0.44$ \\
Bottom friction coefficient $\mu_b$   & $2\times 10^{-5}$ \\
Material density $\rho_0$             & $1060$ kg\,m$^{-3}$ \\
\hline
\end{tabular}
\end{center}
\caption{Microscopic material parameters of the model.}
\label{tab:model_props}
\end{table}

As in the experiment, several packing fractions of the shear-cell are
investigated in the simulations (see Table~\ref{tab:sim_details}).
For too low density, in the {\em sub-critical} regime, the particles
are pushed away from the inner wall and lose contact, so that shearing
stops.  For too high densities, dilation and thus shear are hindered
and the system becomes {\em blocked}.  The intermediate regime $0.793
\le \bar{\nu} \le 0.809$ is of major interest in this study. Note that
the range of densities that allow for the steady state shear flow is
extremely narrow.  However, we remark that the transition points
between the three regimes quantitatively agree between experiments and
simulations.

\begin{table}[tb]
\begin{center}
\begin{tabular}{|c|c|c|c|}
\hline
 Global Volume         & \multicolumn{2}{c|}{Number of Particles} 
                       & Flow Behavior  \\
 Fraction $\bar \nu$   & \ \ small \ \ & large & \\
\hline
\hline
0.789 & 2462 & 404 &\\
0.791 & 2469 & 405 & sub-critical\\
0.793 & 2476 & 406 & ------------ \\
0.796 & 2483 & 407 &\\
0.798 & 2490 & 408 &\\
0.800 & 2498 & 409 &\\
0.800 & 2511 & 400 &\\
0.802 & 2520 & 399 &\\
0.804 & 2511 & 410 & shear flow\\
0.805 & 2524 & 404 &\\
0.807 & 2518 & 412 &\\
0.807 & 2545 & 394 &\\
0.809 & 2525 & 414 & ------------- \\ 
0.810 & 2538 & 407 & ------------- \\
0.811 & 2555 & 399 &\\
0.819 & 2560 & 418 & blocked\\
0.828 & 2588 & 422 &\\
\hline
\end{tabular}
\end{center}
\caption{Details of the simulation runs provided in this study.
Mentioned are those particle numbers for which data were available
in both experiment and simulation.
The horizontal lines in the last column mark the transition between
the sub-critical (the blocked) range of density with the 
shear flow regime.}
\label{tab:sim_details}
\end{table}

Still, there remain some nominally modest differences between the
experiment and the simulation, which may lead to differences between
results for the two realizations. The main differences are:

\begin{itemize}
\item The numerical code used here accounts for a very weak friction with 
  the bottom plate only, presumably smaller than reality and thus allowing 
  less damping of the particles.  In addition, the reduced friction in 
  the experiment is achieved by powder on the bottom plate and this
  may lead to somewhat inhomogeneous friction between the substrate and
  particles.

\item Related to the bottom friction is a possible small tilt of the real 
  particles out of plane of observation, connected to increased
  tangential and frictional forces due to increased, artificial, normal 
  forces.

\item The particle-wall (and also the particle-particle) contacts are
  modeled by simple linear force laws and thus, possibly, do not reproduce
  reality to the extent desired.  More complicated non-linear or hysteretic
  or plastic models 
  \cite{walton86,walton86b,thornton91,thornton97b,mei00,thornton00,thornton00b} are far from the scope of this study.

\item In the original experiment there existed a small bump on the
inner wheel (a deviation from the ring-shape, which in the end leads
to a slightly larger effective radius of the inner wall.  A larger
radius has the strongest effect in the case of low volume fractions,
where the particles are easily moved away from the inner wheel.

\item There is also a difference between the way the initial state is
prepared for the experiments and the simulations.  The starting state
in the experiments is a nearly uniform density at the mean
packing fraction, $\bar{\nu}$.  The initial state of the simulation
is an initially dilated state, which is then compressed.  This
preparation method is described below.

\end{itemize}

These factors apply for all the comparisons between the simulation-
and experimental data to follow. While there are differences in
various details, many qualitative and quantitative results are in
agreement for the experiment and simulation.

\section{Initial conditions and steady state}
\label{sec:initcond}

\begin{figure}
\epsfig{file=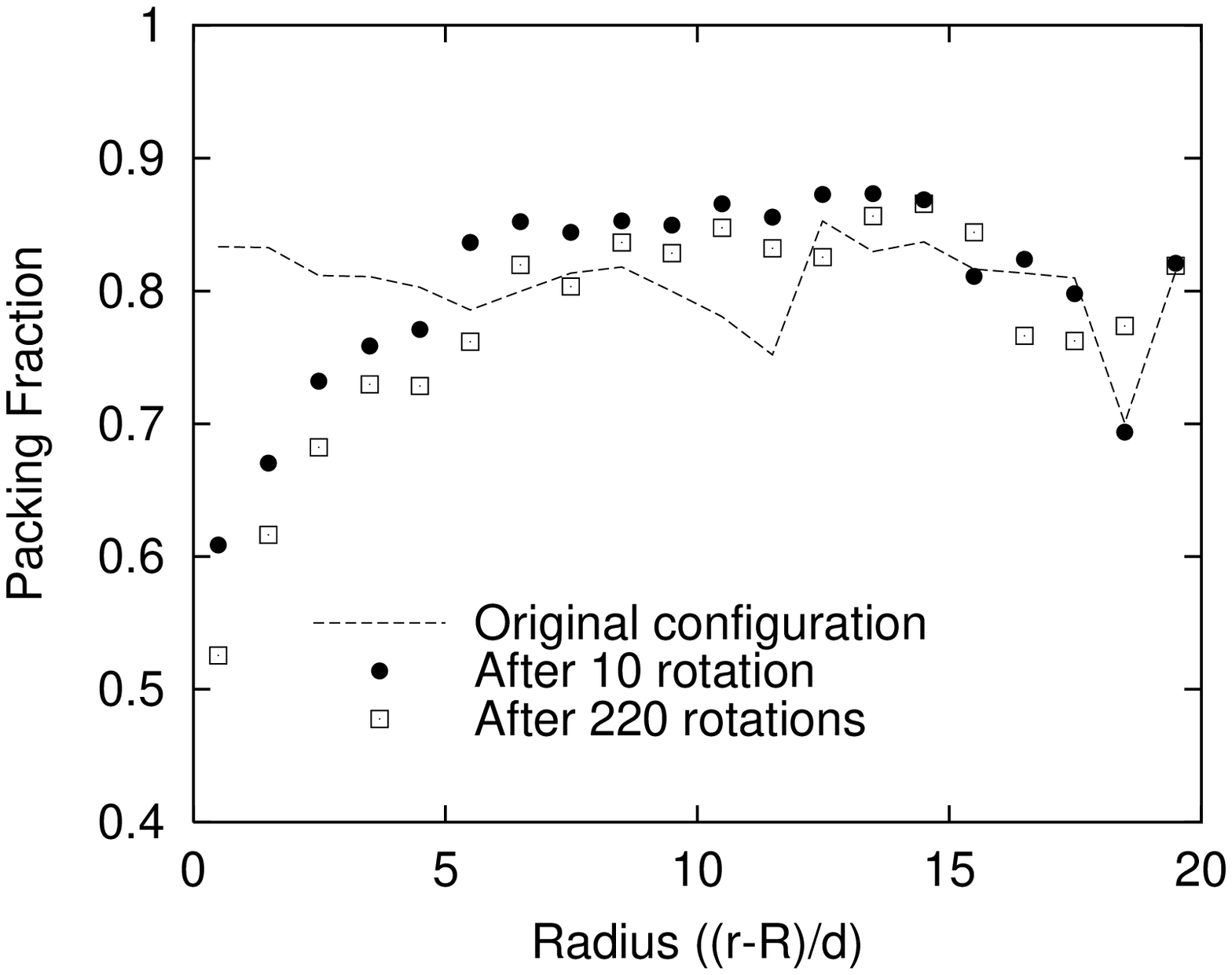,height=7cm,angle=-90}
\epsfig{file=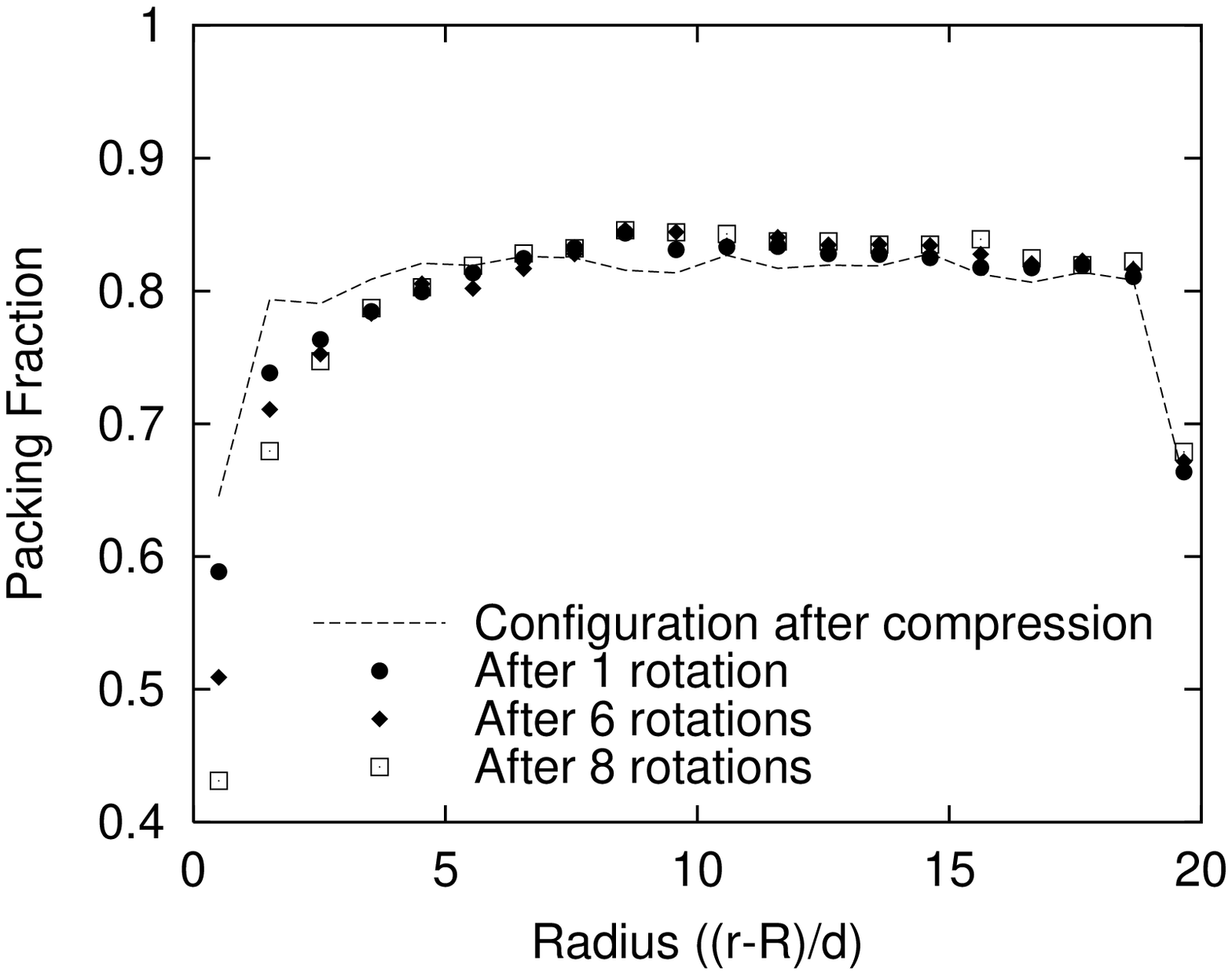,height=7cm,angle=-90}
\caption{
  Evolution of the packing fraction $\nu$ for different times versus
  radial distance $(r-R)/d$ from the wheel in units of disk
  diameters. The upper panel shows experimental data, the lower one
  simulation results.
\label{fig:density}}
\end{figure}

In this section, we explore the initial evolution of the system to a
nominally stationary state, characterized by a dilated region near the
shearing wheel, with large fluctuations in local density and velocity.
Before collecting the data in the experiment, the inner wheel ran
typically for $60\,{\rm min}$ at the highest shearing rate,
corresponding to at least 20 rotations of the inner wheel.  In the
simulation, however, the preparation had to be limited in order to
reduce the comparatively long computation time.  The simulations are
prepared for about three rotation periods, because a few runs with
preparation times of up to ten periods of rotation did {\em not} show
further relaxation effects.  However, the much longer relaxation time
of tens to hundreds of periods as used in the experiment was not
reached, so that long time relaxation effects cannot be ruled out for
the simulations presented here.

\subsection{Preparation Procedure}
\label{sec:prepare}

As noted, the preparation procedure of the simulations is a dilute
state.  Specifically, the system starts with an extended outer ring
$R_{\rm prepare} > R_o=25.24$\,cm.  While the outer ring is expanded,
the inner ring starts to rotate (counterclockwise) with constant
angular velocity $\Omega = 0.1$\,s$^{-1}$.  The radius of the outer
ring is then reduced within about two seconds to reach its desired
value $R_o$. Afterwards, the outer ring is kept fixed and the inner
ring continues to rotate until at $t=t_{\rm max}$ the simulation ends.
Due to the constant volume in the experiment, the disks are inserted
one by one until the desired number and density are reached.  This
difference, which can be seen in Fig.~\ref{fig:density}, affects the
initial density, but should not influence results for the steady
state.

\subsection{Time-evolution of density profiles}

The procedures for establishing a steady state were necessary because
an initial, homogeneous density becomes radially non-uniform as a
consequence of shear-induced dilatancy, for both experiment and
simulation.  Starting from a fairly uniform random packing (dashed
lines in Fig.~\ref{fig:density}, for $\bar{\nu} = 0.804$), a dilated
region forms close to the sheared inner wheel. There are minimal
changes in the density after about 5 rotations of the inner wheel.
Given a CPU-time of 1-2 days per rotation, we did not extend
simulations over more than ten rotations, so that the true long-time
behavior may not be captured here.  Particle rearrangements have been
observed over much longer relaxation times experimentally
\cite{howell99b}.

When making comparisions between the model and the experiment, it is
important to keep in mind the following differences in obtaining local
density data: \\ 

\noindent $\bullet~$ Fill-up procedure (Section \ref{sec:prepare}).
The dashed line from the simulations in Fig.\ \ref{fig:density} is
obtained after a few seconds of compression {\em and} shear, so that a
transient state between the initial and the steady state of the shear
band is visible. The experimental data are obtained from the static
initial state, where no onset of the shear band could take place. \\

\noindent $\bullet~$ Averaging Procedures.  The simulation data are
averaged over full rings around the symmetry center of the shear-cell,
whereas in the experimental system only radial slices that correspond
to one quarter of the entire apparatus were observed. Even though
averages were computed over an extended time interval, a systematic
error due to this procedure cannot be ruled out. Because of possible
circumferential fluctuations associated with this averaging process,
the area under the experimental curves is not necessarily constant,
nor necessarily identical to the global density.\\

\noindent $\bullet~$ 
Experimental density determination.  In the experiment the local
density is measured via optical intensity methods, where there is some
uncertainty due to light scattering and non-linear transmission.\\

Due to these possible systematic differences between the local
densities obtained from experiment and from simulation, we take the
freedom to adjust the local density data, as described below, when
making comparisions between simulations and experiments.

\subsection{Density difference between experiment and simulation}
\label{sec:diffexpsim}

The method used to measure the local packing fraction in the
experiment involves a calibration with some uncertainty, in addition
to the fact that the real particles are not perfect disks as assumed
in the simulation.  Specifically, data are obtained by using the fact
that UV light is strongly attenuated on passing through the
photoelastic disks.  This technique is calibrated against packings
with well known area fractions, such as square and hexagonal lattices.
There are still some small systematic uncertainties in this procedure,
and if one computes the packing fraction using the data given in the
upper part of Figure~\ref{fig:density}, a packing fraction higher than
the global one is found. For that reason, in
Fig.~\ref{fig:rescaleddensity}, we shift the experimental local
density data downward by a constant value of $\nu_{\rm shift}=0.08$.

\section{Changing the Packing Fraction}
\label{sec:diffpack}

In this section, the dependence of the local density, the forces, and
the kinematics of the system are examined as a function of $\bar \nu$,
the mean packing fraction. Using this global density $\bar \nu$ as a
parameter has led to the discovery of a novel transition as the system
approaches a critical packing fraction, $\bar\nu_c$ \cite{howell99}.
In the experiment we found $\bar\nu_c \sim 0.792$ versus $\bar\nu_c
\sim 0.793$ in the simulations.

The reason for this $\bar \nu$-dependence is easy to understand by
imagining what would happen if $\bar \nu$ were very low. In this case,
grains would easily be pushed away from the wheel, and after some
rearrangements they would remain at rest without further contact with
the moving wall.  Increasing $\bar \nu$ by adding more and more grains
would lead to the critical mean density, $\bar\nu_c$, such that there
would always be at least some grains subject to compressive and shear
forces from the boundaries. By adding more grains, the system would
strengthen, more force chains would occur, and grains would be dragged
more frequently by the shearing wheel. If even more particles were
added, the system would become very stiff and eventually would become
blocked, i.e.\ so dense that hardly any shearing can take place. In
the extreme limit, due to large compressive forces and deformations,
permanent plastic deformations might occur and brittle materials even
might fracture.  However, due to the large deformations possible with
polymeric material used in the experiment and due to the relatively
weak forces applied, none of these effects is evidenced.

\subsection{Density}

\begin{figure}
  \begin{center}
    \includegraphics[width=0.7\linewidth,angle=-90]
                    {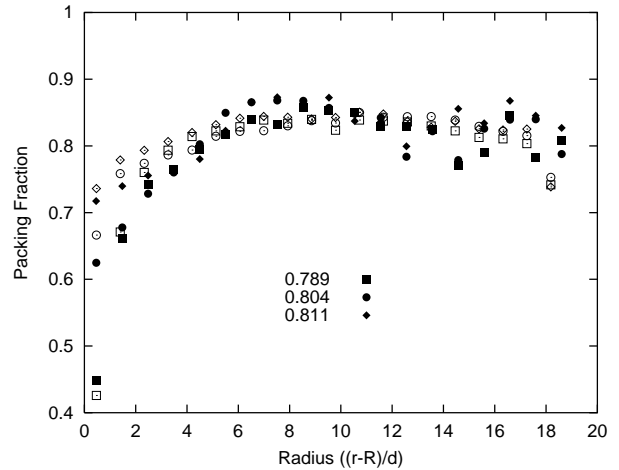}
    
\caption{Volume fraction $\nu$, plotted against the dimensionless
distance from the origin $(r-R)/d$, for different initial global
densities $\bar{\nu}$ (not shifted).  The open symbols give simulation
data with $\bar{\nu}$ as given in the inset.  The solid symbols show
experimental data $\nu-\nu_{\rm shift}$, with $\nu_{\rm
shift}=0.08$. }

\label{fig:rescaleddensity}

\end{center}
\end{figure}

We first consider the local density profiles. In
Fig.~\ref{fig:rescaleddensity}, we show $\nu$ vs.  $(r-R)/d$ for
several $\bar \nu$ values for both experiment and simulation.  The
data show good quantitative agreement within the fluctuations between
experiment and simulation (after the systematic shift-correction
explained above in subsection~\ref{sec:diffexpsim}).  There is a clear
difference in density between the dynamic, dilute shear zone and the
static outer area. From the density data, we infer a width of the
shear zone of about 5-6 particle diameters -- from both experiment and
simulation.

\subsection{Velocity and spin profiles}

\begin{figure}
\epsfig{file=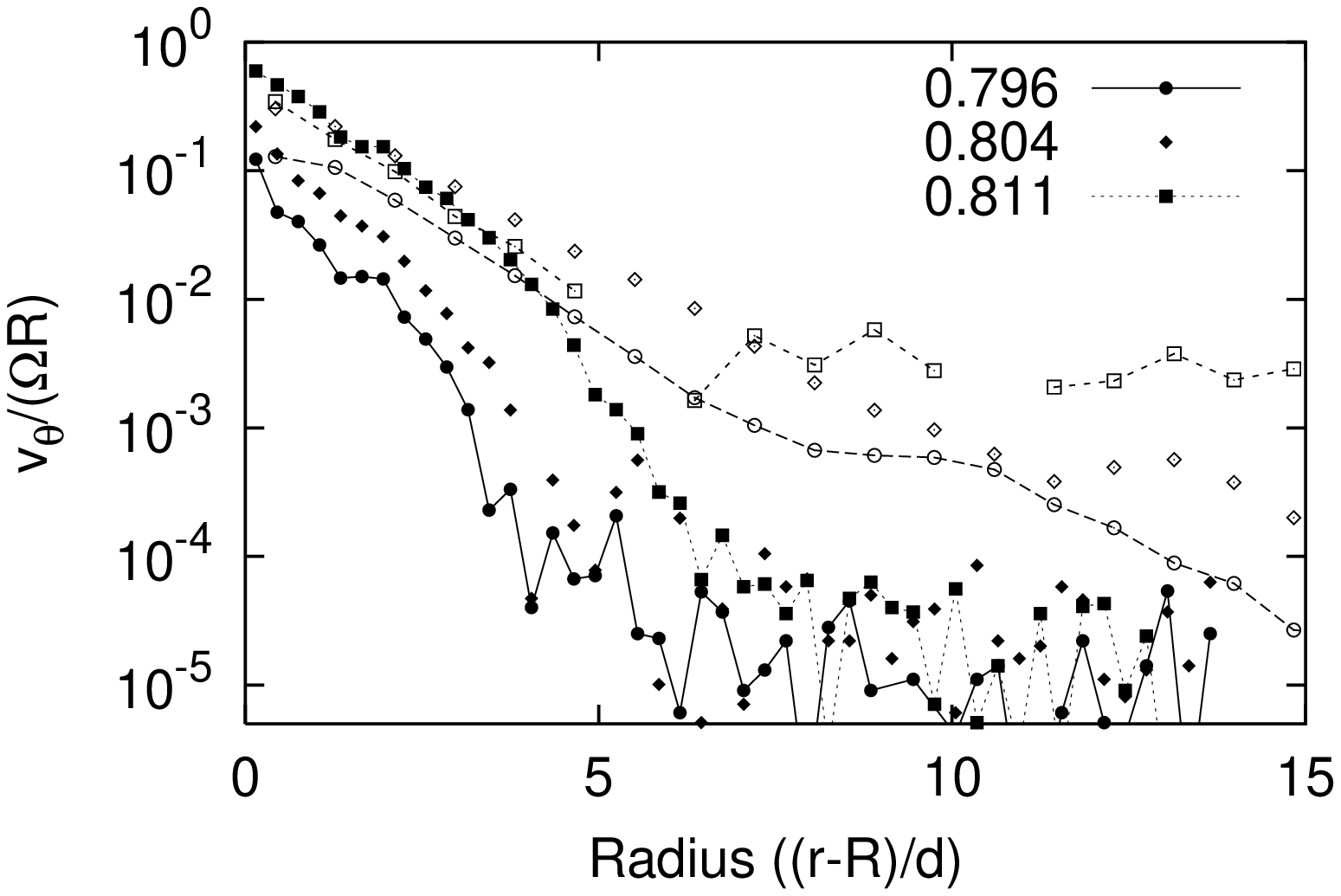,height=0.9\linewidth,angle=-90}
\epsfig{file=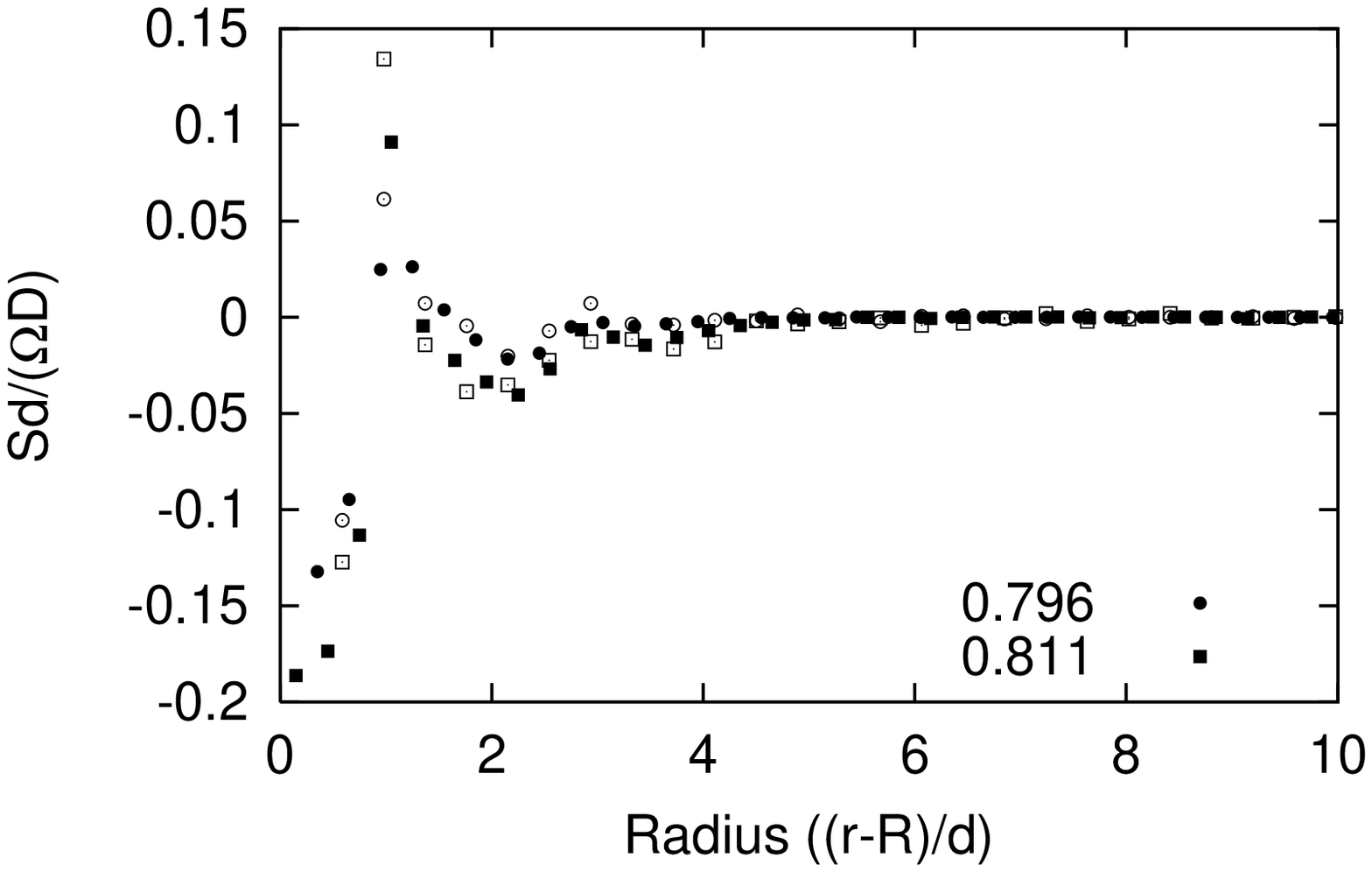,height=0.9\linewidth,angle=-90}
\caption{Velocity and spin profiles for selected packing fractions 
  $\bar \nu$. The solid and open symbols denote experimental and simulational
  data, respectively.
\label{fig:velprofpfexp}}
\end{figure}

In this subsection, we focus on the change in the velocity and spin
profiles with changing $\bar\nu$. In Fig. \ref{fig:velprofpfexp}, we
show data for the velocity profiles for different $\bar\nu$ from both
the experiment and the simulation. The profiles for the normalized
velocity, $v_\theta/(\Omega R)$, show a roughly {\em exponential
decay}, although there is some clear curvature in the experimental
data at the outer edge of the shear zone, where the saturation level
is reached.  This {\em saturation level} of fluctuations in the
velocity is at a higher level in the simulations, possibly due to the
systematically larger shear rate in simulations used to save
CPU-time, or due to the model for bottom friction. However, the 
logarithmic scaling over-amplifies this very small difference.

In the experiment, the amplitude of the exponential term (the velocity
of the particles close to the inner wall, $v_0$) decays steadily to
zero as $\bar \nu$ decreases towards $\bar{\nu}_c$. The simulation
data show a weaker decay of the velocity at the inner wall with
decreasing density.  The fact that $v_0/(\Omega R) \ll 1$ indicates
that as $\nu \rightarrow \nu_c$, either slip or intermittent shear
takes place at the inner wall.  Only values of $v_0/(\Omega R)=1$
would correspond to perfect shear in the sense that the particles are
moving with the wall without slip. For high densities, the agreement
between experiments and simulations is reasonable, but for low
densities, the magnitude of the velocities differs strongly.  This may
be due to either the differences in bottom- or wall-friction, or due
to more irregular and differently shaped walls in the experiments,
causing more intermittency and thus reduced mean velocities.

The experimental and the simulated profiles for the scaled particle
spin, $Sd/(\Omega D)$, evolve in a similar manner with $\bar \nu$.
Oscillations from negative to positive and back to negative spins are
obtained, indicating at least partial rolling of the layers adjacent
to the inner wall \cite{luding96b}.  The mean spins are a little
higher for all the simulations than in the experiment, possibly due to
differences in the bottom friction or due to differences in the
shearing surfaces.

The agreement in the velocity profiles is at least promising, given
that the bottom friction may be wrong by a substantial amount.
Especially for higher densities there is good quantitative agreement.
Indeed, it is for this case that the bottom friction and wall effects
are expected to be
least important, since in this regime, the particle-particle
interaction forces are at their strongest, and intermittent behavior
is much relatively unlikely.

\subsection{Velocity Distributions}

\begin{figure}
  \includegraphics[width=0.6\linewidth,angle=-90,clip=true]{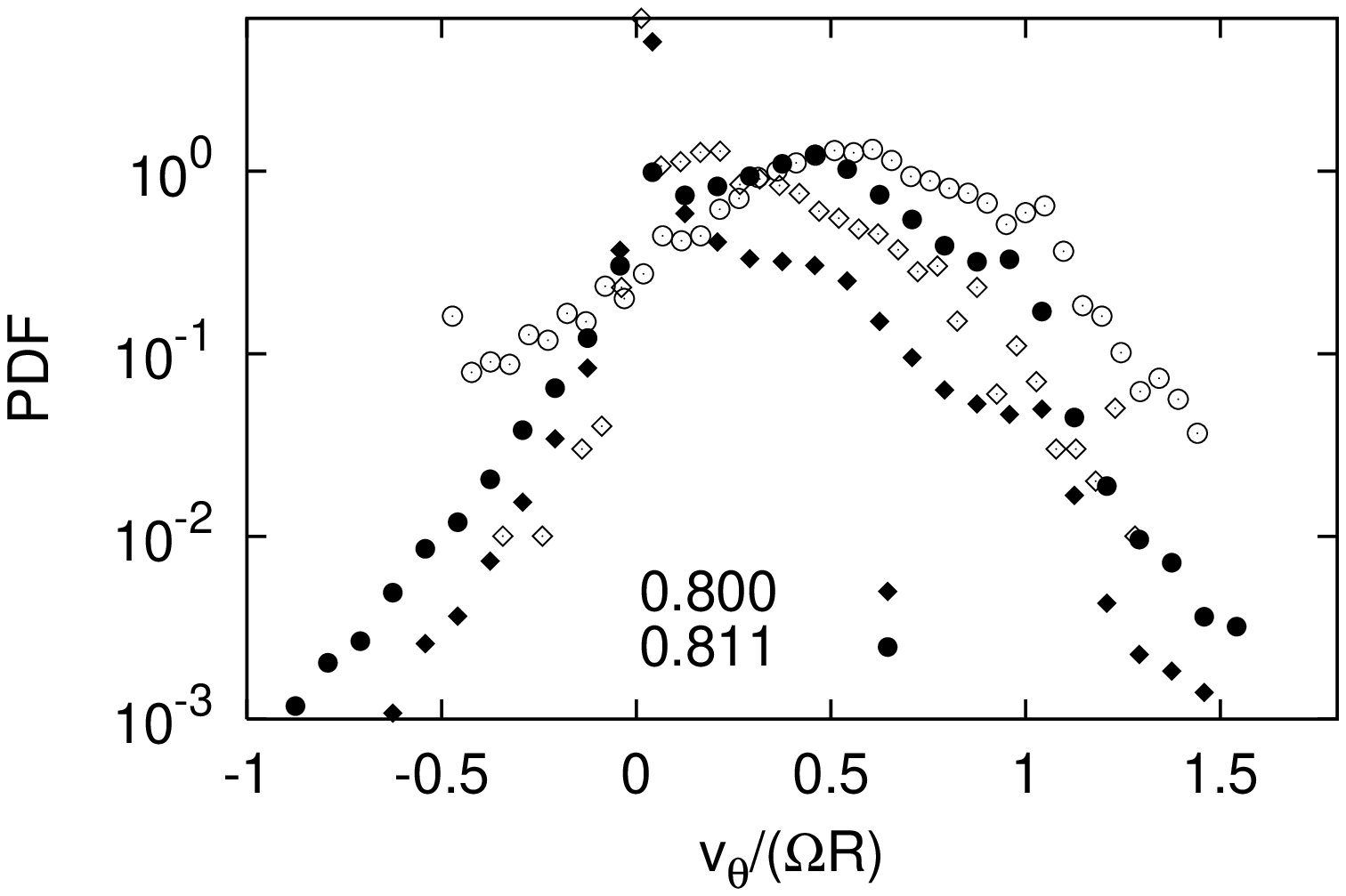}
 \includegraphics[width=0.6\linewidth,angle=-90,clip=true]{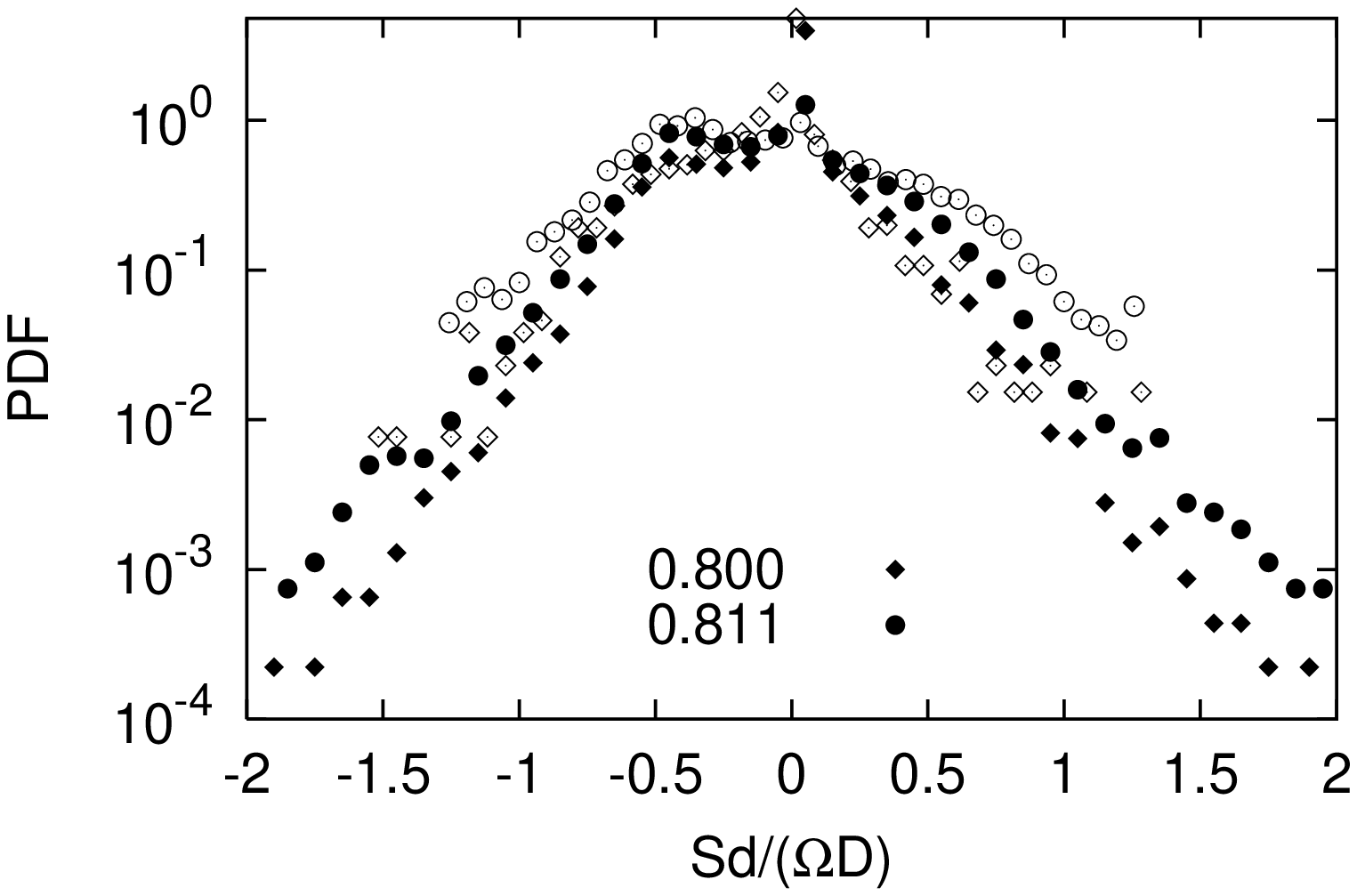}
\caption{Velocity and spin distributions close to the inner wall. 
The solid symbols denote experimental, the open symbols simulation data.
\label{fig:veldistrnu}}
\end{figure}
\begin{figure}
  \psfrag{xlabel}[][][0.6]{$v_\theta/(\Omega R)$}
  \psfrag{ylabel}[][][0.6][90]{$Sd/(\Omega D)$}
  \includegraphics[height=4.2cm,angle=-90]{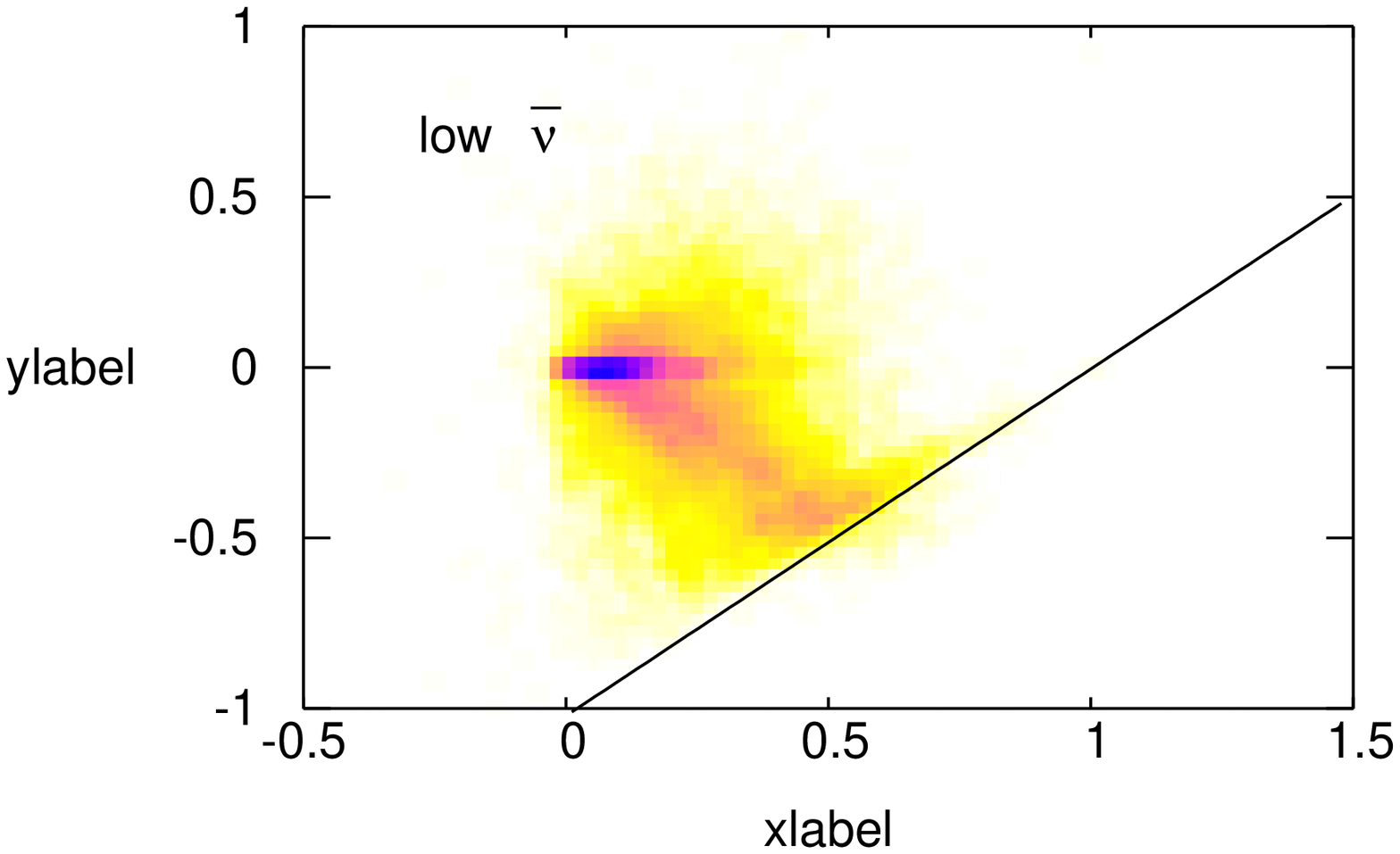}
  \includegraphics[height=4.2cm,angle=-90]{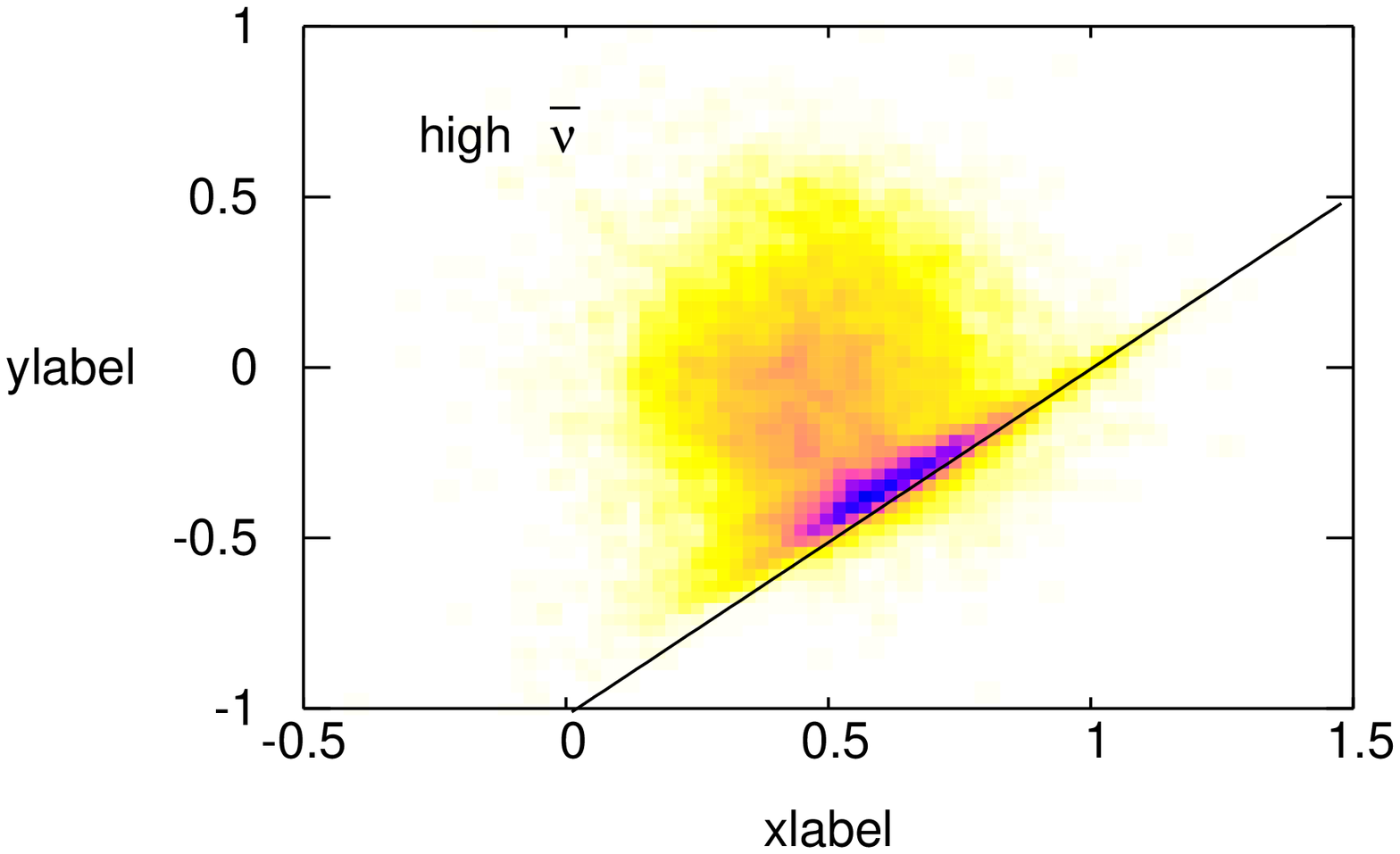}
  \includegraphics[height=4.2cm,angle=-90]{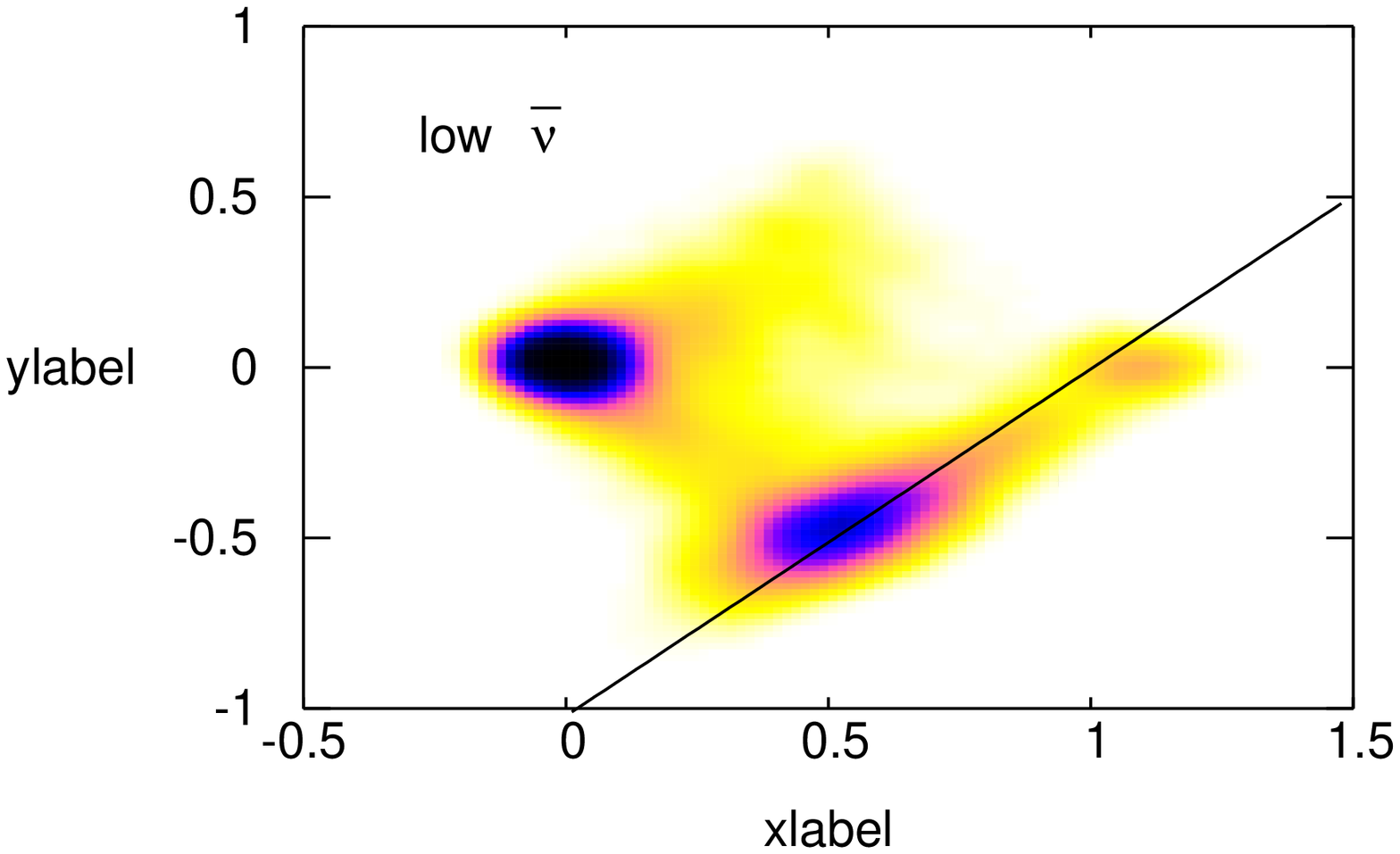}
  \includegraphics[height=4.2cm,angle=-90]{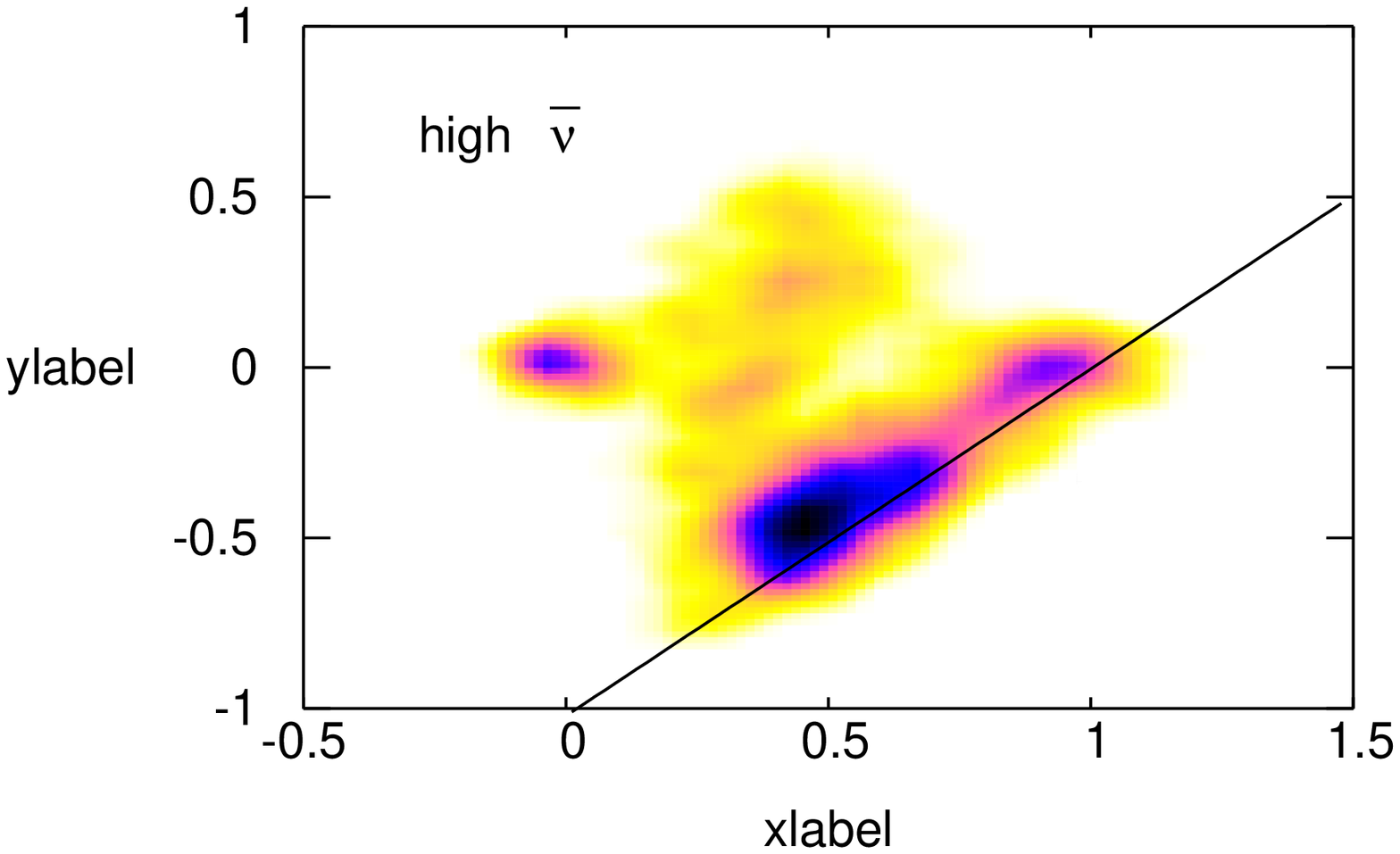}
\caption{2D probability density for $v_\theta/(\Omega R)$ and 
         $Sd/(\Omega D)$ for $0 < (r-R)/d < 1$. Darker regions 
         correspond to higher probability densities. The upper and lower
         panels correspond to simulation and experimental data, 
         respectively.
\label{fig:twodistr}}
\end{figure}

>From the previous section, we infer that changing the packing fraction
affects not only the profiles, but also the distributions of the
velocity. In Fig.~\ref{fig:veldistrnu}, we show the velocity
distributions in a one-particle wide radial bin next to the inner
wheel for various $\bar \nu$-values from the experiment and the
simulation.

The data clearly show that the peaks near the origin, corresponding to
non-rotating particles at rest, become weaker with increasing density.
Furthermore, the regions with negative spin and nonzero $v_\theta$
grow with increasing $\bar \nu$.  The fact that increasing $\bar \nu$
leads to a decreasing number of stationary particles is not
surprising.  But the formation of the second peak in the velocity
distribution at $v_\theta/(\Omega R) \simeq 0.5$ is not as intuitive
as the small peak at unity.

A key to understanding this phenomenon is contained in the
two-variable distribution $P(v_\theta/(\Omega R),Sd/(\Omega D) )$, as
shown in Fig.\ \ref{fig:twodistr} for high (right) and low (left)
density $\bar \nu$. The upper data are experimental and the lower data
are from simulations.  The probability is coded in grayscale with dark
denoting higher probabilities.  This figure indicates two distinct
features, corresponding to two qualitatively different processes. The
first feature is the concentration of probability around $(0,0)$,
which corresponds to a state where the disks are essentially at rest,
without spin or translation. The other feature is a concentration of
probability around the line $v_\theta/(\Omega R)=1+Sd/(\Omega D)$,
which corresponds to non-slip motion of grains relative to the wheel.
No-slip here means, that the particles execute a combination of
backwards rolling and translation, such that the wheel surface and the
disk surface remain in continuous contact.  Thus, the peak at
$v_\theta/(\Omega R)=0$, which is strong for low $\bar \nu$,
corresponds to particles that are so weakly compressed that they can
easily slip with respect to the shearing wheel.  With higher density,
and hence greater force at the contacts between the particles and the
shearing wheel, slipping becomes less likely and the combination of
translation and backwards rolling is the preferred state.

\section{Summary}

We have reported parallel experimental studies and Molecular Dynamics
simulations of shearing in a two-dimensional Couette geometry.  Here,
an important goal was to benchmark such simulations in a setting where
it was possible to have good overlap between the parameters relevant
to the simulations and the experiments.  In most respects, the
numerical results are in good qualitative, and for some quantities
good quantitative agreement with the experimental results
\footnote{This is astonishing when the possible discrepancies
concerning particle shape and boundaries, as well as the partially
huge differences between experimental reality and the
particle-particle and particle-wall contact models in the simulation
are considered}.

Both simulations and experiments show rate-independence within the
statistical errors, and the range of rates that were studied.  We have
particularly focused on the dependence of the shearing states on the
global packing fraction.  Good agreement between simulation and
experiment was found for the density profiles associated with the
formation of a shear band next to the inner shearing wheel with a
characteristic width of about 5 to 6 particle diameters.

Both simulation and experiment also showed a roughly exponential
velocity profile.  However, the simulations did not capture the
density dependence of the experimental velocity profiles, nor some
details of the shape, especially at the outer edge of the shear band.
In this regard, further exploration of the role played by roughness of
the shearing surface and the effect of the particle-bottom friction
are necessary.  The former can lead to more intermittent behavior,
whereas the latter might explain the velocity-drop at the outer edge
of the experimental shear band.

The alternating spin profiles in experiment and simulation agreed
nicely, indicating a rolling of the innermost particle layers (parallel
to the walls) over each other.  Outside of the shear band, rotations
are {\em not} activated, however.  From the velocity- and
spin-probability densities, a combination of rolling and sliding with
the inner wall is evidenced.  As the density decreases towards
$\nu_c$, increasingly more particles remain at rest -- stopped by the
bottom friction.  With increasing density, more and more particles are
dragged with the moving wall, but at the same time roll over each
other -- in layers, with strongly decreasing amplitude as distance
from the moving wall increases.

\section{Conclusion and Outlook}

The present study is of particular interest because of the intensive
attempt to match as many of the detailed properties of the experiment
as feasible by the corresponding properties in the simulation.  
Specifically, most
of the parameters used in the simulation are fixed by experimental
measurement.  Nevertheless, certain properties of the system were
sufficiently complex or difficult to determine exactly, so that there
were some differences between the experimental and simulational
realizations.  In this category of complex properties are friction
with the bottom surface, and the fact that the particles were not
perfectly uniform.  In spite of these differences, all the features
seen in the experiment were also realized in the simulation.  In many
cases, the correspondence between simulation and experiment were
quantitatively correct to within a few percent.  In other cases, the
simulation could be shifted appropriately so that agreement with the
experiment was possible.  Given the uncertainty in experimental
parameters and/or small irregularities, this level of agreement is
quite reasonable.  The clear conclusion is that with sufficient care,
MD modeling of a granular system can produce and predict experimental
behavior, with the understanding that absolute quantitative agreement
may be limited.  Inevitably, in any experiment, small variabilities
between particles or boundaries come into play at a sufficient level
of detail.  At this level agreement between simulation and experiment,
and even between similar but distinct experiments must limit absolute
agreement.


\begin{thebibliography}{45}
\expandafter\ifx\csname natexlab\endcsname\relax\def\natexlab#1{#1}\fi
\expandafter\ifx\csname bibnamefont\endcsname\relax
  \def\bibnamefont#1{#1}\fi
\expandafter\ifx\csname bibfnamefont\endcsname\relax
  \def\bibfnamefont#1{#1}\fi
\expandafter\ifx\csname citenamefont\endcsname\relax
  \def\citenamefont#1{#1}\fi
\expandafter\ifx\csname url\endcsname\relax
  \def\url#1{\texttt{#1}}\fi
\expandafter\ifx\csname urlprefix\endcsname\relax\def\urlprefix{URL }\fi
\providecommand{\bibinfo}[2]{#2}
\providecommand{\eprint}[2][]{\url{#2}}

\bibitem[{\citenamefont{Reynolds}(1885)}]{reynolds85}
\bibinfo{author}{\bibfnamefont{O.}~\bibnamefont{Reynolds}},
  \bibinfo{journal}{Philos. Mag. Ser. 5} \textbf{\bibinfo{volume}{50-20}},
  \bibinfo{pages}{469} (\bibinfo{year}{1885}).

\bibitem[{\citenamefont{Bagnold}(1954)}]{bagnold54}
\bibinfo{author}{\bibfnamefont{R.~A.} \bibnamefont{Bagnold}},
  \bibinfo{journal}{Proc. Royal Soc. London} \textbf{\bibinfo{volume}{225}},
  \bibinfo{pages}{49} (\bibinfo{year}{1954}).

\bibitem[{\citenamefont{Knight}(1997)}]{knight97}
\bibinfo{author}{\bibfnamefont{J.~B.} \bibnamefont{Knight}},
  \bibinfo{journal}{Phys. Rev. E} \textbf{\bibinfo{volume}{55}},
  \bibinfo{pages}{6016} (\bibinfo{year}{1997}).

\bibitem[{\citenamefont{Drake}(1990)}]{drake90}
\bibinfo{author}{\bibfnamefont{T.~G.} \bibnamefont{Drake}},
  \bibinfo{journal}{J. of Geophysical Research} \textbf{\bibinfo{volume}{95}},
  \bibinfo{pages}{8681} (\bibinfo{year}{1990}).

\bibitem[{\citenamefont{Nedderman and Laohakul}(1980)}]{nedderman80}
\bibinfo{author}{\bibfnamefont{R.~M.} \bibnamefont{Nedderman}}
  \bibnamefont{and} \bibinfo{author}{\bibfnamefont{C.}~\bibnamefont{Laohakul}},
  \bibinfo{journal}{Powder Technol.} \textbf{\bibinfo{volume}{25}},
  \bibinfo{pages}{91} (\bibinfo{year}{1980}).

\bibitem[{\citenamefont{Savage and Hutter}(1989)}]{savage89}
\bibinfo{author}{\bibfnamefont{S.~B.} \bibnamefont{Savage}} \bibnamefont{and}
  \bibinfo{author}{\bibfnamefont{K.}~\bibnamefont{Hutter}},
  \bibinfo{journal}{J. Fluid Mech.} \textbf{\bibinfo{volume}{199}},
  \bibinfo{pages}{177} (\bibinfo{year}{1989}).

\bibitem[{\citenamefont{Pouliquen and Gutfraind}(1996)}]{pouliquen96}
\bibinfo{author}{\bibfnamefont{O.}~\bibnamefont{Pouliquen}} \bibnamefont{and}
  \bibinfo{author}{\bibfnamefont{R.}~\bibnamefont{Gutfraind}},
  \bibinfo{journal}{Phys. Rev. E} \textbf{\bibinfo{volume}{53}},
  \bibinfo{pages}{552} (\bibinfo{year}{1996}).

\bibitem[{\citenamefont{Herrmann}(1995)}]{herrmann95}
\bibinfo{author}{\bibfnamefont{H.~J.} \bibnamefont{Herrmann}},
  \bibinfo{journal}{{Physikalische Bl\"atter}} \textbf{\bibinfo{volume}{51}},
  \bibinfo{pages}{1083} (\bibinfo{year}{1995}).

\bibitem[{\citenamefont{Vanel et~al.}(2000)\citenamefont{Vanel, Claudin,
  Bouchaud, Cates, Cl\'ement, and Wittmer}}]{vanel00}
\bibinfo{author}{\bibfnamefont{L.}~\bibnamefont{Vanel}},
  \bibinfo{author}{\bibfnamefont{P.}~\bibnamefont{Claudin}},
  \bibinfo{author}{\bibfnamefont{J.-P.} \bibnamefont{Bouchaud}},
  \bibinfo{author}{\bibfnamefont{M.~E.} \bibnamefont{Cates}},
  \bibinfo{author}{\bibfnamefont{E.}~\bibnamefont{Cl\'ement}},
  \bibnamefont{and} \bibinfo{author}{\bibfnamefont{J.~P.}
  \bibnamefont{Wittmer}}, \bibinfo{journal}{Phys. Rev. Lett.}
  \textbf{\bibinfo{volume}{84}}, \bibinfo{pages}{1439} (\bibinfo{year}{2000}).

\bibitem[{\citenamefont{Geng et~al.}(2001)\citenamefont{Geng, Howell, Longhi,
  Behringer, Reydellet, Vanel, Cl\'ement, and Luding}}]{geng01}
\bibinfo{author}{\bibfnamefont{J.}~\bibnamefont{Geng}},
  \bibinfo{author}{\bibfnamefont{D.}~\bibnamefont{Howell}},
  \bibinfo{author}{\bibfnamefont{E.}~\bibnamefont{Longhi}},
  \bibinfo{author}{\bibfnamefont{R.~P.} \bibnamefont{Behringer}},
  \bibinfo{author}{\bibfnamefont{G.}~\bibnamefont{Reydellet}},
  \bibinfo{author}{\bibfnamefont{L.}~\bibnamefont{Vanel}},
  \bibinfo{author}{\bibfnamefont{E.}~\bibnamefont{Cl\'ement}},
  \bibnamefont{and} \bibinfo{author}{\bibfnamefont{S.}~\bibnamefont{Luding}},
  \bibinfo{journal}{Phys. Rev. Lett.} \textbf{\bibinfo{volume}{87}},
  \bibinfo{pages}{035506} (\bibinfo{year}{2001}).

\bibitem[{\citenamefont{Geng et~al.}(2002)\citenamefont{Geng, Behringer,
  Reydellet, Cl\'ement, and Luding}}]{geng02}
\bibinfo{author}{\bibfnamefont{J.}~\bibnamefont{Geng}},
  \bibinfo{author}{\bibfnamefont{R.~P.} \bibnamefont{Behringer}},
  \bibinfo{author}{\bibfnamefont{G.}~\bibnamefont{Reydellet}},
  \bibinfo{author}{\bibfnamefont{E.}~\bibnamefont{Cl\'ement}},
  \bibnamefont{and} \bibinfo{author}{\bibfnamefont{S.}~\bibnamefont{Luding}}
  (\bibinfo{year}{2002}), \bibinfo{note}{to be published}.

\bibitem[{\citenamefont{Veje et~al.}(1999)\citenamefont{Veje, Howell, and
  Behringer}}]{veje99}
\bibinfo{author}{\bibfnamefont{C.~T.} \bibnamefont{Veje}},
  \bibinfo{author}{\bibfnamefont{D.~W.} \bibnamefont{Howell}},
  \bibnamefont{and} \bibinfo{author}{\bibfnamefont{R.~P.}
  \bibnamefont{Behringer}}, \bibinfo{journal}{Phys. Rev. E}
  \textbf{\bibinfo{volume}{59}}, \bibinfo{pages}{739} (\bibinfo{year}{1999}).

\bibitem[{\citenamefont{Miller et~al.}(1996)\citenamefont{Miller, O'{H}ern, and
  Behringer}}]{miller96}
\bibinfo{author}{\bibfnamefont{B.}~\bibnamefont{Miller}},
  \bibinfo{author}{\bibfnamefont{C.}~\bibnamefont{O'{H}ern}}, \bibnamefont{and}
  \bibinfo{author}{\bibfnamefont{R.~P.} \bibnamefont{Behringer}},
  \bibinfo{journal}{Phys. Rev. Lett.} \textbf{\bibinfo{volume}{77}},
  \bibinfo{pages}{3110} (\bibinfo{year}{1996}).

\bibitem[{\citenamefont{Veje et~al.}(1998)\citenamefont{Veje, Howell,
  Behringer, Sch\"ollmann, Luding, and Herrmann}}]{veje98}
\bibinfo{author}{\bibfnamefont{C.~T.} \bibnamefont{Veje}},
  \bibinfo{author}{\bibfnamefont{D.~W.} \bibnamefont{Howell}},
  \bibinfo{author}{\bibfnamefont{R.~P.} \bibnamefont{Behringer}},
  \bibinfo{author}{\bibfnamefont{S.}~\bibnamefont{Sch\"ollmann}},
  \bibinfo{author}{\bibfnamefont{S.}~\bibnamefont{Luding}}, \bibnamefont{and}
  \bibinfo{author}{\bibfnamefont{H.~J.} \bibnamefont{Herrmann}}, in
  \emph{\bibinfo{booktitle}{Physics of Dry Granular Media}}, edited by
  \bibinfo{editor}{\bibfnamefont{H.~J.} \bibnamefont{Herrmann}},
  \bibinfo{editor}{\bibfnamefont{J.-P.} \bibnamefont{Hovi}}, \bibnamefont{and}
  \bibinfo{editor}{\bibfnamefont{S.}~\bibnamefont{Luding}}
  (\bibinfo{publisher}{Kluwer Academic Publishers},
  \bibinfo{address}{Dordrecht}, \bibinfo{year}{1998}), p. \bibinfo{pages}{237}.

\bibitem[{\citenamefont{Howell et~al.}(1999{\natexlab{a}})\citenamefont{Howell,
  Behringer, and Veje}}]{howell99}
\bibinfo{author}{\bibfnamefont{D.}~\bibnamefont{Howell}},
  \bibinfo{author}{\bibfnamefont{R.~P.} \bibnamefont{Behringer}},
  \bibnamefont{and} \bibinfo{author}{\bibfnamefont{C.}~\bibnamefont{Veje}},
  \bibinfo{journal}{Phys. Rev. Lett.} \textbf{\bibinfo{volume}{82}},
  \bibinfo{pages}{5241} (\bibinfo{year}{1999}{\natexlab{a}}).

\bibitem[{\citenamefont{Howell et~al.}(1999{\natexlab{b}})\citenamefont{Howell,
  Behringer, and Veje}}]{howell99b}
\bibinfo{author}{\bibfnamefont{D.~W.} \bibnamefont{Howell}},
  \bibinfo{author}{\bibfnamefont{R.~P.} \bibnamefont{Behringer}},
  \bibnamefont{and} \bibinfo{author}{\bibfnamefont{C.~T.} \bibnamefont{Veje}},
  \bibinfo{journal}{Chaos} \textbf{\bibinfo{volume}{9}}, \bibinfo{pages}{559}
  (\bibinfo{year}{1999}{\natexlab{b}}).

\bibitem[{\citenamefont{Howell}(1999)}]{howell99c}
\bibinfo{author}{\bibfnamefont{D.~W.} \bibnamefont{Howell}}, Ph.D. thesis,
  \bibinfo{school}{Duke University, Department of Physics}
  (\bibinfo{year}{1999}).

\bibitem[{\citenamefont{Azanza et~al.}(1999)\citenamefont{Azanza, chevoir, and
  Moucheront}}]{azanza99}
\bibinfo{author}{\bibfnamefont{E.}~\bibnamefont{Azanza}},
  \bibinfo{author}{\bibfnamefont{F.}~\bibnamefont{chevoir}}, \bibnamefont{and}
  \bibinfo{author}{\bibfnamefont{P.}~\bibnamefont{Moucheront}},
  \bibinfo{journal}{J. Fluid Mech.} \textbf{\bibinfo{volume}{400}},
  \bibinfo{pages}{199} (\bibinfo{year}{1999}).

\bibitem[{\citenamefont{Losert et~al.}(1999)\citenamefont{Losert, Cooper, and
  Gollub}}]{losert99}
\bibinfo{author}{\bibfnamefont{W.}~\bibnamefont{Losert}},
  \bibinfo{author}{\bibfnamefont{D.~G.~W.} \bibnamefont{Cooper}},
  \bibnamefont{and} \bibinfo{author}{\bibfnamefont{J.~P.}
  \bibnamefont{Gollub}}, \bibinfo{journal}{Phys. Rev. E}
  \textbf{\bibinfo{volume}{59}}, \bibinfo{pages}{5855} (\bibinfo{year}{1999}).

\bibitem[{\citenamefont{Buggisch and L\"offelmann}(1989)}]{buggisch89}
\bibinfo{author}{\bibfnamefont{H.}~\bibnamefont{Buggisch}} \bibnamefont{and}
  \bibinfo{author}{\bibfnamefont{G.}~\bibnamefont{L\"offelmann}},
  \bibinfo{journal}{Chemical Engineering and Processing}
  \textbf{\bibinfo{volume}{26}}, \bibinfo{pages}{193} (\bibinfo{year}{1989}).

\bibitem[{\citenamefont{Rotter et~al.}(1998)\citenamefont{Rotter, Holst, Ooi,
  and Sanad}}]{rotter98}
\bibinfo{author}{\bibfnamefont{J.~M.} \bibnamefont{Rotter}},
  \bibinfo{author}{\bibfnamefont{J.~M. F.~G.} \bibnamefont{Holst}},
  \bibinfo{author}{\bibfnamefont{J.~Y.} \bibnamefont{Ooi}}, \bibnamefont{and}
  \bibinfo{author}{\bibfnamefont{A.~M.} \bibnamefont{Sanad}},
  \bibinfo{journal}{Phil. Trans. R. Soc. Lond. A}
  \textbf{\bibinfo{volume}{356}}, \bibinfo{pages}{2685} (\bibinfo{year}{1998}).

\bibitem[{\citenamefont{Holst et~al.}(1999{\natexlab{a}})\citenamefont{Holst,
  Ooi, Rotter, and Rong}}]{holst99}
\bibinfo{author}{\bibfnamefont{J.~M. F.~G.} \bibnamefont{Holst}},
  \bibinfo{author}{\bibfnamefont{J.~Y.} \bibnamefont{Ooi}},
  \bibinfo{author}{\bibfnamefont{J.~M.} \bibnamefont{Rotter}},
  \bibnamefont{and} \bibinfo{author}{\bibfnamefont{G.~H.} \bibnamefont{Rong}},
  \bibinfo{journal}{Journal of Engineering Mechanics ASCE}
  \textbf{\bibinfo{volume}{125}}, \bibinfo{pages}{94}
  (\bibinfo{year}{1999}{\natexlab{a}}).

\bibitem[{\citenamefont{Holst et~al.}(1999{\natexlab{b}})\citenamefont{Holst,
  Rotter, Ooi, and Rong}}]{holst99b}
\bibinfo{author}{\bibfnamefont{J.~M. F.~G.} \bibnamefont{Holst}},
  \bibinfo{author}{\bibfnamefont{J.~M.} \bibnamefont{Rotter}},
  \bibinfo{author}{\bibfnamefont{J.~Y.} \bibnamefont{Ooi}}, \bibnamefont{and}
  \bibinfo{author}{\bibfnamefont{G.~H.} \bibnamefont{Rong}},
  \bibinfo{journal}{Journal of Engineering Mechanics ASCE}
  \textbf{\bibinfo{volume}{125}}, \bibinfo{pages}{104}
  (\bibinfo{year}{1999}{\natexlab{b}}).

\bibitem[{\citenamefont{Luding et~al.}(2001{\natexlab{a}})\citenamefont{Luding,
  L\"atzel, and Herrmann}}]{luding01}
\bibinfo{author}{\bibfnamefont{S.}~\bibnamefont{Luding}},
  \bibinfo{author}{\bibfnamefont{M.}~\bibnamefont{L\"atzel}}, \bibnamefont{and}
  \bibinfo{author}{\bibfnamefont{H.~J.} \bibnamefont{Herrmann}}, in
  \emph{\bibinfo{booktitle}{Handbook of Conveying and Handling of Particulate
  Solids}}, edited by \bibinfo{editor}{\bibfnamefont{A.}~\bibnamefont{Levy}}
  \bibnamefont{and} \bibinfo{editor}{\bibfnamefont{H.}~\bibnamefont{Kalman}}
  (\bibinfo{publisher}{Elsevier}, \bibinfo{address}{Amsterdam, The
  Netherlands}, \bibinfo{year}{2001}{\natexlab{a}}), pp.
  \bibinfo{pages}{39--44}.

\bibitem[{\citenamefont{Luding et~al.}(2001{\natexlab{b}})\citenamefont{Luding,
  L\"atzel, Volk, Diebels, and Herrmann}}]{luding01b}
\bibinfo{author}{\bibfnamefont{S.}~\bibnamefont{Luding}},
  \bibinfo{author}{\bibfnamefont{M.}~\bibnamefont{L\"atzel}},
  \bibinfo{author}{\bibfnamefont{W.}~\bibnamefont{Volk}},
  \bibinfo{author}{\bibfnamefont{S.}~\bibnamefont{Diebels}}, \bibnamefont{and}
  \bibinfo{author}{\bibfnamefont{H.~J.} \bibnamefont{Herrmann}},
  \bibinfo{journal}{Comp. Meth. Appl. Mech. Engng.}
  \textbf{\bibinfo{volume}{191}}, \bibinfo{pages}{21}
  (\bibinfo{year}{2001}{\natexlab{b}}).

\bibitem[{\citenamefont{Campbell and Brennen}(1985)}]{campbell85}
\bibinfo{author}{\bibfnamefont{C.~S.} \bibnamefont{Campbell}} \bibnamefont{and}
  \bibinfo{author}{\bibfnamefont{C.~E.} \bibnamefont{Brennen}},
  \bibinfo{journal}{J. Fluid. Mech.} \textbf{\bibinfo{volume}{151}},
  \bibinfo{pages}{167} (\bibinfo{year}{1985}).

\bibitem[{\citenamefont{Evans and Morriss}(1983)}]{evans83}
\bibinfo{author}{\bibfnamefont{D.~J.} \bibnamefont{Evans}} \bibnamefont{and}
  \bibinfo{author}{\bibfnamefont{G.~P.} \bibnamefont{Morriss}},
  \bibinfo{journal}{Phys. Rev. Lett.} \textbf{\bibinfo{volume}{51}},
  \bibinfo{pages}{1776} (\bibinfo{year}{1983}).

\bibitem[{\citenamefont{Walton and Braun}(1986{\natexlab{a}})}]{walton86b}
\bibinfo{author}{\bibfnamefont{O.~R.} \bibnamefont{Walton}} \bibnamefont{and}
  \bibinfo{author}{\bibfnamefont{R.~L.} \bibnamefont{Braun}},
  \bibinfo{journal}{Acta Mechanica} \textbf{\bibinfo{volume}{63}},
  \bibinfo{pages}{73} (\bibinfo{year}{1986}{\natexlab{a}}).

\bibitem[{\citenamefont{Thornton and Yin}(1991)}]{thornton91}
\bibinfo{author}{\bibfnamefont{C.}~\bibnamefont{Thornton}} \bibnamefont{and}
  \bibinfo{author}{\bibfnamefont{K.~K.} \bibnamefont{Yin}},
  \bibinfo{journal}{Powder Technol.} \textbf{\bibinfo{volume}{65}},
  \bibinfo{pages}{153} (\bibinfo{year}{1991}).

\bibitem[{\citenamefont{Thornton}(1997)}]{thornton97b}
\bibinfo{author}{\bibfnamefont{C.}~\bibnamefont{Thornton}},
  \bibinfo{journal}{Journal of Applied Mechanics}
  \textbf{\bibinfo{volume}{64}}, \bibinfo{pages}{383} (\bibinfo{year}{1997}).

\bibitem[{\citenamefont{Thornton and Antony}(2000)}]{thornton00b}
\bibinfo{author}{\bibfnamefont{C.}~\bibnamefont{Thornton}} \bibnamefont{and}
  \bibinfo{author}{\bibfnamefont{S.~J.} \bibnamefont{Antony}},
  \bibinfo{journal}{Powder Technology} \textbf{\bibinfo{volume}{109}},
  \bibinfo{pages}{179} (\bibinfo{year}{2000}).

\bibitem[{\citenamefont{L\"atzel et~al.}(2000)\citenamefont{L\"atzel, Luding,
  and Herrmann}}]{latzel00}
\bibinfo{author}{\bibfnamefont{M.}~\bibnamefont{L\"atzel}},
  \bibinfo{author}{\bibfnamefont{S.}~\bibnamefont{Luding}}, \bibnamefont{and}
  \bibinfo{author}{\bibfnamefont{H.~J.} \bibnamefont{Herrmann}},
  \bibinfo{journal}{Granular Matter} \textbf{\bibinfo{volume}{2}},
  \bibinfo{pages}{123} (\bibinfo{year}{2000}),
  \bibinfo{note}{cond-mat/0003180}.

\bibitem[{\citenamefont{Masson}(2001)}]{masson01}
\bibinfo{author}{\bibfnamefont{S.}~\bibnamefont{Masson}}, \bibinfo{journal}{J.
  Engineering Mechanics} \textbf{\bibinfo{volume}{127}}, \bibinfo{pages}{1007}
  (\bibinfo{year}{2001}).

\bibitem[{\citenamefont{Tillemans and Herrmann}(1995)}]{tillemans95}
\bibinfo{author}{\bibfnamefont{H.-J.} \bibnamefont{Tillemans}}
  \bibnamefont{and} \bibinfo{author}{\bibfnamefont{H.~J.}
  \bibnamefont{Herrmann}}, \bibinfo{journal}{Physica A}
  \textbf{\bibinfo{volume}{217}}, \bibinfo{pages}{261} (\bibinfo{year}{1995}).

\bibitem[{\citenamefont{Thornton and Zhang}(2001)}]{thornton01}
\bibinfo{author}{\bibfnamefont{C.}~\bibnamefont{Thornton}} \bibnamefont{and}
  \bibinfo{author}{\bibfnamefont{L.}~\bibnamefont{Zhang}}, in
  \emph{\bibinfo{booktitle}{Powders \& Grains 2001}}, edited by
  \bibinfo{editor}{\bibfnamefont{Y.}~\bibnamefont{Kishino}}
  (\bibinfo{publisher}{Balkema}, \bibinfo{address}{Rotterdam},
  \bibinfo{year}{2001}), pp. \bibinfo{pages}{183--190}.

\bibitem[{\citenamefont{Oda and Iwashita}(2000)}]{oda00}
\bibinfo{author}{\bibfnamefont{M.}~\bibnamefont{Oda}} \bibnamefont{and}
  \bibinfo{author}{\bibfnamefont{K.}~\bibnamefont{Iwashita}},
  \bibinfo{journal}{Int. J. of Enginering Science}
  \textbf{\bibinfo{volume}{38}}, \bibinfo{pages}{1713} (\bibinfo{year}{2000}).

\bibitem[{\citenamefont{Oda and Kazama}(1998)}]{oda98}
\bibinfo{author}{\bibfnamefont{M.}~\bibnamefont{Oda}} \bibnamefont{and}
  \bibinfo{author}{\bibfnamefont{H.}~\bibnamefont{Kazama}},
  \bibinfo{journal}{G\'eotechnique} \textbf{\bibinfo{volume}{48}},
  \bibinfo{pages}{465} (\bibinfo{year}{1998}).

\bibitem[{\citenamefont{Radjai et~al.}(1999)\citenamefont{Radjai, Roux, and
  Moreau}}]{radjai99}
\bibinfo{author}{\bibfnamefont{F.}~\bibnamefont{Radjai}},
  \bibinfo{author}{\bibfnamefont{S.}~\bibnamefont{Roux}}, \bibnamefont{and}
  \bibinfo{author}{\bibfnamefont{J.~J.} \bibnamefont{Moreau}},
  \bibinfo{journal}{Chaos} \textbf{\bibinfo{volume}{9}}, \bibinfo{pages}{544}
  (\bibinfo{year}{1999}).

\bibitem[{\citenamefont{Sch\"ollmann}(1999)}]{schollmann99}
\bibinfo{author}{\bibfnamefont{S.}~\bibnamefont{Sch\"ollmann}},
  \bibinfo{journal}{Phys. Rev. E} \textbf{\bibinfo{volume}{59}},
  \bibinfo{pages}{889} (\bibinfo{year}{1999}).

\bibitem[{\citenamefont{Luding}(2002)}]{luding02b}
\bibinfo{author}{\bibfnamefont{S.}~\bibnamefont{Luding}},
  \bibinfo{journal}{Advances in Complex Systems} \textbf{\bibinfo{volume}{4}},
  \bibinfo{pages}{379} (\bibinfo{year}{2002}).

\bibitem[{\citenamefont{L\"atzel}(1999)}]{latzel99}
\bibinfo{author}{\bibfnamefont{M.}~\bibnamefont{L\"atzel}}, Master's thesis,
  \bibinfo{school}{Universit{\"a}t Stuttgart} (\bibinfo{year}{1999}).

\bibitem[{\citenamefont{Walton and Braun}(1986{\natexlab{b}})}]{walton86}
\bibinfo{author}{\bibfnamefont{O.~R.} \bibnamefont{Walton}} \bibnamefont{and}
  \bibinfo{author}{\bibfnamefont{R.~L.} \bibnamefont{Braun}},
  \bibinfo{journal}{Journal of Rheology} \textbf{\bibinfo{volume}{30}},
  \bibinfo{pages}{949} (\bibinfo{year}{1986}{\natexlab{b}}).

\bibitem[{\citenamefont{Mei et~al.}(2000)\citenamefont{Mei, Shang, Walton, and
  Klausner}}]{mei00}
\bibinfo{author}{\bibfnamefont{R.~W.} \bibnamefont{Mei}},
  \bibinfo{author}{\bibfnamefont{H.}~\bibnamefont{Shang}},
  \bibinfo{author}{\bibfnamefont{O.~R.} \bibnamefont{Walton}},
  \bibnamefont{and} \bibinfo{author}{\bibfnamefont{J.~F.}
  \bibnamefont{Klausner}}, \bibinfo{journal}{Powder Technology}
  \textbf{\bibinfo{volume}{112}}, \bibinfo{pages}{102} (\bibinfo{year}{2000}).

\bibitem[{\citenamefont{Thornton}(2000)}]{thornton00}
\bibinfo{author}{\bibfnamefont{C.}~\bibnamefont{Thornton}},
  \bibinfo{journal}{G\'eotechnique} \textbf{\bibinfo{volume}{50}},
  \bibinfo{pages}{43} (\bibinfo{year}{2000}).

\bibitem[{\citenamefont{Luding et~al.}(1996)\citenamefont{Luding, Duran,
  Cl\'ement, and Rajchenbach}}]{luding96b}
\bibinfo{author}{\bibfnamefont{S.}~\bibnamefont{Luding}},
  \bibinfo{author}{\bibfnamefont{J.}~\bibnamefont{Duran}},
  \bibinfo{author}{\bibfnamefont{E.}~\bibnamefont{Cl\'ement}},
  \bibnamefont{and}
  \bibinfo{author}{\bibfnamefont{J.}~\bibnamefont{Rajchenbach}},
  \bibinfo{journal}{J. Phys. I France} \textbf{\bibinfo{volume}{6}},
  \bibinfo{pages}{823} (\bibinfo{year}{1996}).

\end{thebibliography}

\end{document}